\documentclass[pra,twocolumn,nofootinbib,floatfix]{revtex4-1} % subeqn
\usepackage[usenames]{color}
\usepackage{hyperref}

\usepackage{bbold}

\usepackage{hhline}

\usepackage[utf8]{inputenc}
\usepackage[T1]{fontenc}
\usepackage[pdftex]{graphicx}

\usepackage{amsmath, amsxtra, amssymb, mathtools, isomath, xspace} %, icomma, stmaryrd, nccmath
\renewcommand{\vec}{\vectorsym}
\usepackage{siunitx}

\usepackage[british]{babel}

\usepackage{booktabs}
\usepackage{multirow}
\usepackage{sidecap}

\newcommand{\mycite}[1]{\cite{#1}}

\newcommand{\ket}[1]{\ensuremath{\vert #1 \rangle}\xspace}%
\newcommand{\bra}[1]{\ensuremath{\langle #1 \vert}\xspace}%
\newcommand{\abs}[1]{\ensuremath{\lvert #1 \rvert}\xspace}%

\newcommand{\Rbk}{\ensuremath{R_{\text b}}\xspace}%
\newcommand{\Nat}{\ensuremath{{N_{\text{at}}}}\xspace}%
\newcommand{\Ne}{\ensuremath{{N_{\text{e}}}}\xspace}%
\newcommand{\alat}{\ensuremath{a_{\text{lat}}}\xspace}%

\long\def\symbolfootnote[#1]#2{\begingroup%
\def\thefootnote{\fnsymbol{footnote}}\footnotetext[#1]{#2}\endgroup}

\sidecaptionvpos{figure}{c}

\begin{document}

% \title{\hspace*{-1cm} \begin{minipage}[b]{1.1\linewidth}\centering \bf Observation of mesoscopic crystalline structures\\ in a two-dimensional Rydberg gas \end{minipage}}
\title{Dynamical crystallization in a low-dimensional Rydberg gas}

%% Notice placement of commas and superscripts and use of &
%% in the author listx

\author{Peter~Schau\ss$^{1,*}$}%
\author{Johannes~Zeiher$^{1}$}%
\author{Takeshi Fukuhara$^{1}$}%
\author{Sebastian Hild$^{1}$}%
\author{Marc Cheneau$^{2}$}%
\author{Tommaso~Macr\`i$^{3}$}%
\author{Thomas Pohl$^{3}$}%
\author{Immanuel Bloch$^{1,4}$}%
\author{Christian Gross$^{1}$}%

\date{3 April 2014}

%\vspace{0.2cm} 
\affiliation{$^1$Max-Planck-Institut f\"{u}r Quantenoptik, 85748 Garching, Germany}
\affiliation{$^2$Laboratoire Charles Fabry, Institut d'optique Graduate School - CNRS - Universit\'e Paris Sud, 91127 Palaiseau, France}%
\affiliation{$^3$Max-Planck-Institut f\"ur Physik komplexer Systeme, 01187 Dresden, Germany}%
\affiliation{$^4$Ludwig-Maximilians-Universit\"{a}t, Fakult\"{a}t f\"{u}r Physik, 80799 M\"{u}nchen, Germany}%

%}

\begin{abstract}
Dominating finite-range interactions in many-body systems can lead to
  intriguing self-ordered phases of matter. Well known examples are
  crystalline solids or Coulomb crystals in ion traps. In those systems, crystallization proceeds via a classical transition, driven by
  thermal fluctuations. In contrast, ensembles of ultracold atoms laser-excited
  to Rydberg states provide a well-controlled quantum system~\mycite{Saffman2010}, in which a
  crystalline phase transition governed by quantum fluctuations can be explored~\mycite{Pohl2010a,Schachenmayer2010,VanBijnen2011}. Here
  we report on the experimental preparation of the crystalline states in such a
  Rydberg many-body system. Fast coherent control on the many-body level is
  achieved via numerically optimized laser excitation
  pulses~\mycite{Pohl2010a,Schachenmayer2010,VanBijnen2011}. We observe an excitation-number
  staircase~\mycite{Weimer2008,Pohl2010a,VanBijnen2011,Schachenmayer2010,Weimer2010,Sela2011}
  as a function of the system size and show directly the emergence of
  incompressible ordered states on its steps. Our results demonstrate
  the applicability of quantum optical control techniques in strongly
  interacting systems, paving the way towards the investigation of novel
  quantum phases in long-range interacting quantum systems, as well as for
  detailed studies of their coherence and correlation
  properties~\mycite{Weimer2008,Pohl2010a,VanBijnen2011,Schachenmayer2010,Weimer2010,Sela2011,Lesanovsky2011}.
\end{abstract}

  \symbolfootnote[1]{Electronic address: {\bf peter.schauss@mpq.mpg.de}}
  \maketitle%

%%%%%%%%%%%%%%%%%%%%%%%%%%%%%%%%%%%%%%%%%%%%%%%%%%%%%%%%%%%%%%%%%%%%%%%%%%%
%                                  Intro                                  %
%%%%%%%%%%%%%%%%%%%%%%%%%%%%%%%%%%%%%%%%%%%%%%%%%%%%%%%%%%%%%%%%%%%%%%%%%%%
% crystallization of Photons (Lukin?)

% First sentence is true for atoms and ions
Rydberg atoms exhibit unique properties that are key to realize and explore
novel quantum many-body Hamiltonians and their phases. 
% Now atoms, tailored interactions
The  strong van der Waals interaction between them allows to create
many-body systems with tailored long-range interactions in neutral ultra-cold
atom samples~\mycite{Saffman2010,Low2012,Hofmann2013a}. 
% optical excitation
Complete experimental control of these systems is possible using the well
developed toolbox of quantum optics for the laser-excitation to the Rydberg
states.
% properties of the interactions
The magnitude of the resulting interactions between the Rydberg atoms is
determined by the choice of the excited state and it can exceed all other
relevant energy scales on distances of several microns, thereby leading to
an ensemble dominated by long-range interactions. 
% effect of the interactions in many body systems 
In this regime, the ground state of the resulting many-body system is expected
to show crystalline ordering of the Rydberg excitations, which can be understood
in the limit of vanishing coupling as the classical closest packing of hard
spheres~\mycite{Ji2011a}.  
% Blockade sets the size. use bandwidth as a word that fits to sweep, rabi and
% line width
The lattice constant of the crystal is set by the dipole blockade radius \Rbk
~\mycite{Jaksch2000, Lukin2001}, defined as the inter-particle
spacing at which the dipole interaction between two Rydberg atoms exceeds the
spectral range of the optical coupling. 
% Preparation
To prepare the system in this crystalline phase, a dynamical approach has been
suggested that adiabatically connects the ground state containing no Rydberg
excitations with the targeted crystalline state. At the heart of this
\textit{dynamical crystallization} technique is the coherent control of the
many-body system~\mycite{Pohl2010a, Schachenmayer2010,VanBijnen2011,Beterov2011a,Brierley2012,Petrosyan2013a,Ebert2014}.

Previous experiments showed direct or indirect evidence for 
correlations caused by the long-range interactions in Rydberg many-body systems,
such as a universal scaling of the Rydberg excitation
number~\mycite{Loew2009}, sub-Poissonian counting
statistics~\mycite{Malossi2013,Schempp2014,Ebert2014}, photon correlations~\mycite{Dudin2012b,Peyronel2012}, characteristic exciton dynamics~\mycite{Bettelli2013} or spatial ordering of the
excitations~\mycite{Schwarzkopf2011,Viteau2011a,Schauss2012}.
However, all measurements have predominantly probed features of excited many-body states
that -- next to the ground state -- are also strongly influenced by
the dramatically enhanced interaction scales.

%%%%%%%%%%%%%%%%%%%%%%%%%%%%%%%%%%%%%%%%%%%%%%%%%%%%%%%%%%%%%%%%%%%%%%%%%%%
%                             Here we show ..                             %
%%%%%%%%%%%%%%%%%%%%%%%%%%%%%%%%%%%%%%%%%%%%%%%%%%%%%%%%%%%%%%%%%%%%%%%%%%%

% ------- Figure 1 -------
\begin{figure*}
 
\centering \includegraphics{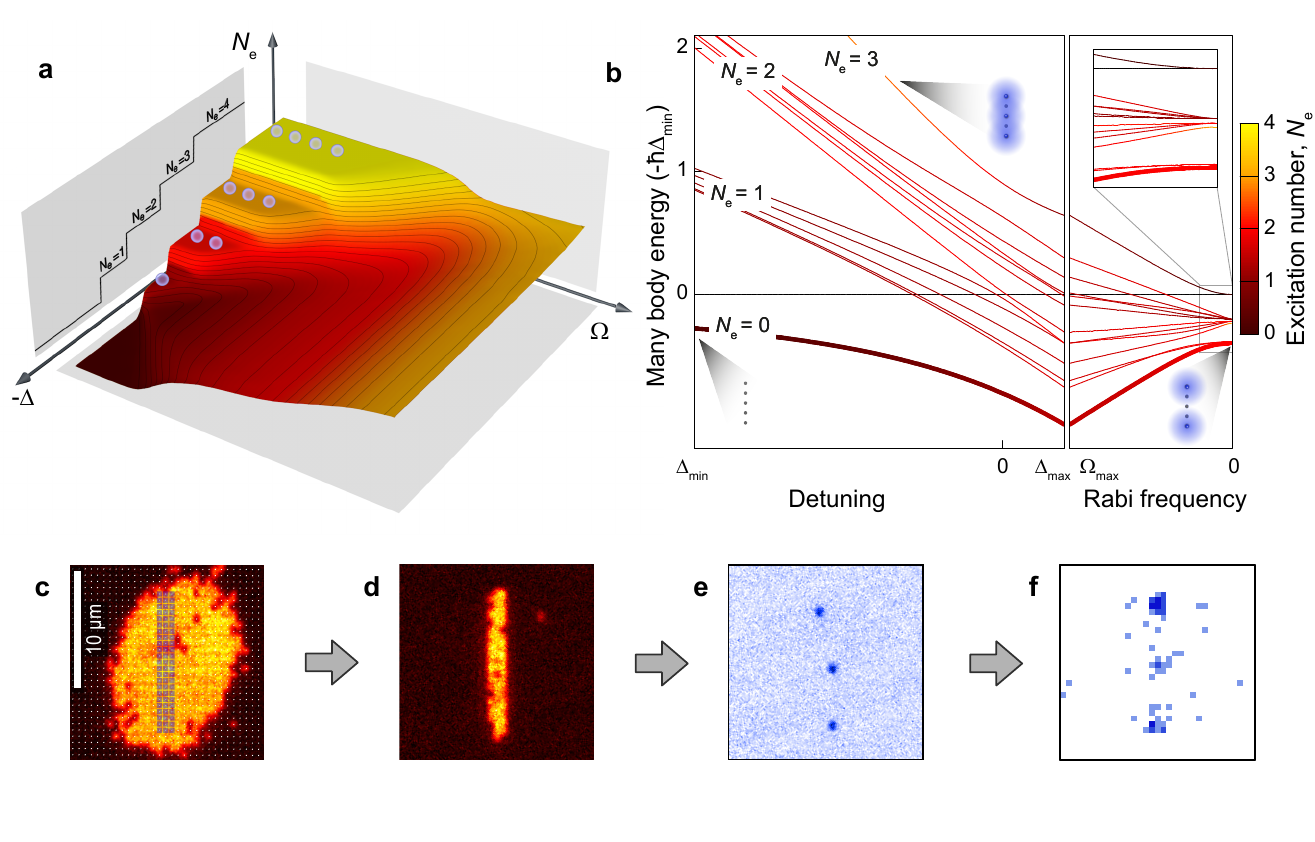} %[width=\textwidth]
\caption{{\bf Schematic illustrating the phase diagram, energy spectrum and
experimental sequence.}  \textbf{a}, Schematic phase diagram. The colour
scale indicates the number of Rydberg excitations in the many-body ground
state of a one-dimensional system. The number of Rydberg excitations \Ne is
visualized in the crystalline phase by the small spheres. For illustration,
the detuning $\Delta$ and Rabi frequency $\Omega$ axes are rescaled by their
sixth root. \textbf{b}, Schematic illustrating the evolution of the excitation
spectrum during a sweep, where coupling strength $\Omega(t)$ and detuning
$\Delta(t)$ are controlled. The spectrum was calculated for an exemplary 1D
system of five atoms. First, the detuning is changed from $\Delta_\text{min}$
to $\Delta_\text{max}$ at constant Rabi frequency $\Omega_{\rm max}$, with
$\Delta_{\rm max}$ chosen to prepare $\Ne=2$ Rydberg excitations.
Subsequently, the Rabi frequency is reduced from $\Omega_\text{max}$ to $0$.
The inset is a zoom into the end of the sweep highlighting the shrinking
gap between the energy levels. The colour of each line indicates the
mean number of Rydberg excitations in the many-body state. For strongly
negative detuning the four different manifolds correspond to the excitation
number Fock states, whose occupation \Ne is indicated in the figure.
In three limiting cases in which the states become classical the spatial excitation pattern is shown 
(blue circles: Rydberg atoms, grey circles: ground state atoms).
\textbf{c-e}, Exemplary fluorescence pictures from different times in the
experimental cycle. \textbf{c}, Mott insulator with lattice sites (white
dots) and spatial light modulator pattern (semi-transparent overlay).
\textbf{d}, Initial atom configuration, \textbf{e}, Single shot Rydberg
excitation pattern. \textbf{f}, Rydberg excitation density after averaging $40$ experimental runs (darker colour means more detected atoms).
}\label{fig:1} 

\end{figure*}
  
Here, we report on the deterministic preparation of crystalline many-body
states in small Rydberg systems via coherently controlled excitation as
proposed in refs.~\mycite{Pohl2010a, Schachenmayer2010,VanBijnen2011}. For one-dimensional
systems we experimentally realize the dipole-blockade staircase and
demonstrate the emergence of crystalline states with a vanishing
compressibility. We trace the evolution of the quantum state during the
dynamic preparation by spatially resolved detection of the excitations at
successive times. Additionally, we developed a novel technique to control the
initial atomic density distribution (Methods). This has turned out to be a crucial
requirement to observe low energy states, especially for small two-dimensional
systems where the many-body energy spectrum is significantly denser than in the
one-dimensional case and strongly depends on the initial shape. In such
two-dimensional systems we observe a sharp concentration of the Rydberg
excitations along the edge of the initial cloud, which is expected for the
energetically low lying many-body states.

%%%%%%%%%%%%%%%%%%%%%%%%%%%%%%%%%%%%%%%%%%%%%%%%%%%%%%%%%%%%%%%%%%%%%%%%%%%
%          Introduction to the physics of our system                      %
%%%%%%%%%%%%%%%%%%%%%%%%%%%%%%%%%%%%%%%%%%%%%%%%%%%%%%%%%%%%%%%%%%%%%%%%%%%

The physical system studied here is a well-defined line- or disc-shaped atomic
sample in an optical lattice with one atom per site. The rubidium-$87$ atoms
are coupled to the Rydberg state 43S$_{1/2}$ with a controlled time-dependent
Rabi frequency $\Omega(t)$ and detuning $\Delta(t)$. The corresponding
theoretical model describing this system is the so-called "frozen Rydberg gas"
Hamiltonian, in which only the internal electronic degrees of freedom are
considered. This is justified by the short time scale on which our experiments
take place during which the motion of the atoms in the lattice is
negligible~\mycite{Anderson1998,Mourachko1998}. Adopting a two-level description, the many-body dynamics of the internal atomic states is governed by the Hamiltonian

\begin{widetext}
\begin{equation}\label{eq:1}
\begin{split} \hat{H} = \frac{\hbar\Omega(t)}{2} \sum_{\vec i}
  \left(\ket{e^{(\vec i)}}\bra{g^{(\vec i)}} + \ket{g^{(\vec i)}}\bra{e^{(\vec i)}}\right) + \sum_{\vec i \neq
  \vec j} \frac{V_{\vec i \vec j}}{2} \hat n^{(\vec i)}_{e} \hat n^{(\vec
  j)}_{e} -  \hbar \Delta(t) \sum_i \hat n^{(\vec i)}_{e}\; . 
\end{split}
\end{equation} 
\end{widetext}
Here, the vectors $\vec i = (i_{x}, i_{y})$ label the position of the atoms on
the lattice. The operators $\ket{e^{(\vec i)}}\bra{g^{(\vec i)}}$ and
$\ket{g^{(\vec i)}}\bra{e^{(\vec i)}}$ describe the transition from the ground state
$\ket{g}$ to the Rydberg state $\ket{e}$, while the operator $\hat n_e^{(\vec
i)} = \ket{e^{(\vec i)}}\bra{e^{(\vec i)}}$ represents the Rydberg state occupation on site $(\vec i)$. The first term
of the Hamiltonian describes the coherent coupling between ground and excited
state with the time-dependent Rabi frequency $\Omega(t)$. The second term
arises due to the van der Waals interaction between two atoms in the Rydberg
state. For the 43S$_{1/2}$ state of rubidium-$87$ $V_{\vec i\vec j} =
-C_{6}/r_{\vec i\vec j}^{6}$ is repulsive due to a negative van der Waals
coefficient $C_{6}<0$. Here, $r_{\vec i\vec j} = \alat |\vec i - \vec j|$ is
the distance between two atoms on the lattice with period $\alat$. The last
term corresponds to an effective chemical potential for the Rydberg excitations that can be
controlled by the detuning $\Delta(t)$.

%%%%%%%%%%%%%%%%%%%%%%%%%%%%%%%%%%%%%%%%%%%%%%%%%%%%%%%%%%%%%%%%%%%%%%%%%%%
%          Fig1ab: Details of the Frozen gas Hamiltonian                    %
%%%%%%%%%%%%%%%%%%%%%%%%%%%%%%%%%%%%%%%%%%%%%%%%%%%%%%%%%%%%%%%%%%%%%%%%%%%

% ------- Figure 2 -------
\begin{SCfigure*}
  \caption{{\bf Identification of the crystalline phase.} \textbf{a}, Excitation
number versus system length for a one-dimensional system. Blue circles
correspond to the experimental mean number of Rydberg excitations \Ne after the optimized sweep. The right axis shows the excitation
number corrected for the detection efficiency. The green line is the result
of the numerical simulation for the experimental initial states, taking into account an initial state filling of $0.8$ and length fluctuations of the order of one site. The grey line shows the classical
($\Omega=0$) prediction. Insets: Measured spatial distribution of Rydberg
excitations (left) and corresponding theory (right) for system lengths $\ell$ of $12$, $23$ and $35$ sites. The
brightness (light to dark) translates to the normalized number of
excitations. \textbf{b}, Compressibility $\kappa$ of the prepared states. Blue circles are 
derived from the experimental data shown in \textbf{a} using a numerical derivative.
The green line is a direct numerical result (Methods).
 All error bars s.e.m.}\label{fig:2} 
\centering \includegraphics{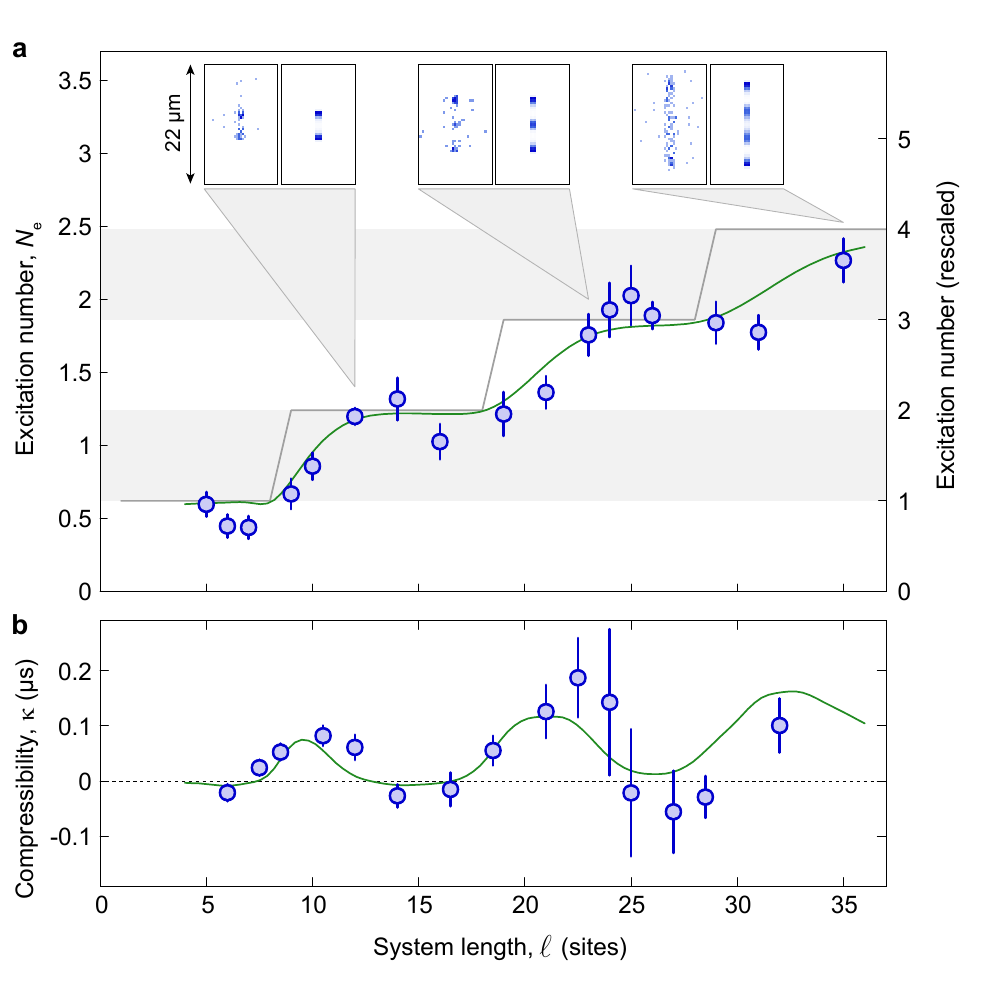} %[width=\textwidth]
\end{SCfigure*}
  
%%Static properties of Ground State  
Rydberg atoms have recently been discussed as a platform for the simulation of
quantum magnetism. Especially, the frozen gas Hamiltonian has been at the focus
of theoretical and experimental interest due to its rich variety of
strongly-correlated phases~\mycite{Weimer2008,Weimer2010,Sela2011}. Introducing spin-$1/2$
operators, the Hamiltonian can be rewritten in the form of an Ising model with
long-range spin interactions in an effective transverse ($\hbar \Omega$) and longitudinal
($-\hbar \Delta$) magnetic field~\mycite{Weimer2010, Lesanovsky2011}. In the classical limit,
$\Omega=0$ and for $\Delta>0$, the many-body ground state corresponds to
crystalline Fock states with a total excitation number $\Ne =\langle
\hat{\Ne}\rangle=\sum_{\vec i} \langle \hat{n}^{(\vec i)}_e\rangle$.
Consequently, the Rydberg excitation number \Ne forms a complete devil's
staircase~\mycite{Bak1982} as a function of $\Delta$ in the thermodynamic limit. In a
one-dimensional chain of $\ell\gg \Ne$ lattice sites, the excitation number
increases from $\Ne$ to $\Ne+1$ at the critical detunings 
$\ell^6\hbar\Delta_{\rm c}\approx 7 \abs{C_6} \Ne^6 / \alat^6$ separating successive
crystal states with a lattice spacing $\alat\ell/(\Ne-1)$ \mycite{Pohl2010a}. The laser coupling
introduces quantum fluctuations, whose effect has been studied in a number of
recent theory works~\mycite{Weimer2008,Weimer2010,Sela2011,Lesanovsky2011,Ates2012b}. Upon
increasing $\Omega$, it has been predicted that, in the thermodynamic limit,
the system undergoes a two-stage quantum melting~\mycite{Weimer2010,Sela2011}
via an incommensurate floating solid with algebraic correlations followed by a
Kosterlitz-Thouless transition~\mycite{Weimer2010,Sela2011} to a disordered phase. The
corresponding scenario for a finite lattice is shown schematically in
Fig.~\ref{fig:1}a. While finite size effects naturally broaden the transitions
in the $(\Omega,\Delta)$ parameter space, extended lobes corresponding to
crystalline states of \Ne excitations with vanishing number fluctuations can
be well identified for typical parameters of our experiments.

%%Preparation of the crystalline phase
The preparation of the crystalline states requires a fast dynamical control due
to the short lifetime of the Rydberg states of typically several tens of microseconds. Our
initial state with all atoms in their electronic ground state coincides with the
many-body ground state of the system for negative detunings and $\Omega=0$.
Since for small coupling strength $\Omega$ the energy gap to the first excited
state closes at the transition points $\Delta_{\rm c}$ between successive
\Ne-manifolds, $\Omega$ and $\Delta$ have to be varied simultaneously in
order to maximize the adiabaticity of the preparation scheme. An intuitive and
simple choice of the path $(\Omega(t), \Delta(t))$ starts with a large negative
detuning $\Delta_{\rm min}$ at which the coupling $\Omega$ is switched
on~\mycite{Pohl2010a,Schachenmayer2010,VanBijnen2011}. Next, the detuning is increased to the
desired final blue-detuned value $\Delta_{\rm max}>0$, followed by a gradual reduction of the coupling strength $\Omega$ to zero.
 In the final stage of this last
step the energy of several many-body states becomes nearly degenerate, as illustrated 
in Fig.~\ref{fig:1}b for an exemplary system of five excitations. These
lowest many-body excited states all belong to the same \Ne-manifold but feature a finite density of dislocations with respect to the perfectly ordered classical ground state. In practice this leads to unavoidable
non-adiabatic transitions at the end of the laser pulse, resulting in
non-classical crystalline states composed of spatially localized collective
excitations~\mycite{Pohl2010a}.

%%%%%%%%%%%%%%%%%%%%%%%%%%%%%%%%%%%%%%%%%%%%%%%%%%%%%%%%%%%%%%%%%%%%%%%%%%%
%          Fig1c: Description of the experimental sequence                %
%%%%%%%%%%%%%%%%%%%%%%%%%%%%%%%%%%%%%%%%%%%%%%%%%%%%%%%%%%%%%%%%%%%%%%%%%%%

% ------- Figure 3 -------
\begin{SCfigure*}

\centering
\includegraphics{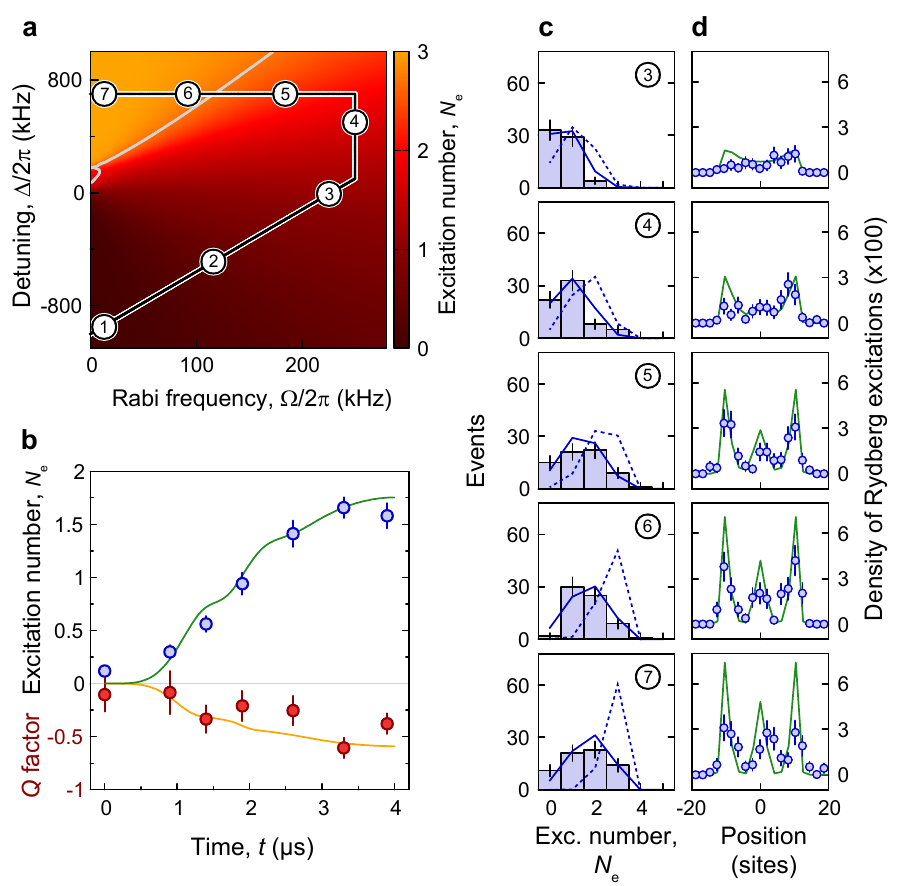}%[width=\textwidth]
\caption{{\bf Dynamical crystallization.}  \textbf{a}, Illustration of the
laser sweep. The black line shows the path of the sweep through the phase
diagram, the numbered positions mark the measurements (c.f. \textbf{c}). The grey
line indicates the boundary of the crystalline lobes, where the $Q$ factor drops below $-0.9$. The phase diagram was calculated for the
experimental parameters. \textbf{b}, Mean number of Rydberg excitations (blue
circles) and $Q$ factor (red circles) for the seven times marked in \textbf{a}
together with the theoretical prediction. \textbf{c}, Experimental and
theoretical probability distributions of the number of Rydberg excitations
along the sweep (c.f. \textbf{a}). Blue boxes show experimental data and the dashed
and solid lines represent the theoretical result for detection
efficiencies of $\alpha=1$ and $\alpha=0.62$, respectively. \textbf{d},
Transversally averaged distributions (probability per site) of the
excitations for the same times as in \textbf{c} with a binning of two sites (blue
circles). The slight asymmetry towards the right might be due to a gradient
in the Rabi frequency (Methods). The green line is the numerical result. All error bars
s.e.m. }\label{fig:3}
% references to other subfigures are also bold without braces (checked in printed nature).

\end{SCfigure*}

Our experiment started from a two-dimensional degenerate gas of approximately
\numrange{250}{700} rubidium-$87$ atoms confined to a single antinode of a
vertical ($z$-axis) optical lattice. The gas was driven deep into the
Mott-insulating phase by adiabatically turning on a square optical lattice with
period $\alat = \SI{532}{nm}$ in the $xy$-plane~\mycite{Sherson2010}. We
used a deconfining beam to reduce the harmonic potential induced by the lattice
beams and thereby enlarged the spatial extension of a single occupancy Mott
insulating state~\mycite{Fukuhara2013b}. Next, we prepared the
initial atomic density distribution precisely by cutting out the desired cloud
shape from the initial Mott insulator using a spatial light modulator
(Fig.~\ref{fig:1}c-d and Methods). For our measurements we chose line- or
disc-shaped atomic samples of well controlled length or radius. The line had a
width of three lattice sites and a variable length $\ell$. Since this width was
much smaller than the blockade radius of approximately nine sites, this
geometry realized an effective one-dimensional chain with a collectively
enhanced Rabi frequency $\sqrt{3}\Omega$. The average filling was
$0.8\,$atoms/site and at the edge it dropped to below $0.1\,$atoms/site, within
one lattice site. The coupling to the Rydberg state was realized by a
two-photon process via the intermediate state $5P_{3/2}$,
using laser wavelengths of \SIlist{780 ; 480}{nm} with $\sigma^{-}$ and
$\sigma^{+}$ polarizations, respectively~\mycite{Schauss2012}. Detailed
excitation beam parameters are summarized in Table 1. Fast
control of the Rabi frequency $\Omega(t)$ and the detuning $\Delta(t)$ was
implemented by tuning intensity and frequency of the $\SI{780}{nm}$ excitation laser using a calibrated
acousto-optical modulator (Methods). Finally, the Rydberg atoms were
detected locally by fluorescence imaging after removing the ground state atoms
from the trap and de-pumping the Rydberg state back to the ground state (Methods)~\mycite{Schauss2012}.
 The spatial distribution of Rydberg atoms was
measured by averaging over at least $40$ realizations (Fig.~\ref{fig:1}e-f). It
directly reveals the crystalline order in small one-dimensional systems, where
the translational invariance is broken and all crystalline excitation patterns
with $\Ne>1$ are pinned to the boundary.

%%%%%%%%%%%%%%%%%%%%%%%%%%%%%%%%%%%%%%%%%%%%%%%%%%%%%%%%%%%%%%%%%%%%%%%%%%%
%          Fig2: Staircase in 1d                                          %
%%%%%%%%%%%%%%%%%%%%%%%%%%%%%%%%%%%%%%%%%%%%%%%%%%%%%%%%%%%%%%%%%%%%%%%%%%%

In a first series of experiments, we prepared crystalline states in the 1D
geometry. For fixed system size the weak scaling of the blockade radius $\Rbk$
with the detuning $\Rbk \propto \Delta_\text{max}^{-1/6}$ limits the maximal excitation
number that can be realized experimentally. Hence, instead of varying the
detuning, we changed the length $\ell$ of the initial system to explore the
characteristics of the Rydberg crystals~\mycite{Pohl2010a}. We measured the
mean number of Rydberg excitations \Ne for varying length
$\ell$ using a numerically optimized sweep (Fig.~\ref{edfig:1}a and Methods). 
In the optimization the sweep duration was set to \SI{4}{\micro\second}, which is a reasonable compromise between the decreasing detection efficiency for longer sweeps and adiabaticity % the increasing adiabaticity for longer sweeps
 (Methods). The results for the sweep to $\Delta_{\rm max}=2\pi \cdot
700(200)\,$kHz shown in Fig.~\ref{fig:2}a exhibit clear plateaus in the mean
excitation number \Ne and agree well with numerical
predictions which take into account the initial atomic density, the actual sweep and the detection
efficiency $\alpha=0.62(5)$ (Methods). On the plateaus the theory predicts strong overlap with Fock states (Fig.~\ref{edfig:3}). Exploiting the fact that varying the detuning
$\Delta_{\rm max}$ is approximately equivalent to varying the system size $\ell$ (Methods),
 we extract the compressibility $\kappa=\frac{\partial \Ne}{\partial
\Delta_{\rm max}}$ from our data. As shown in Fig.~\ref{fig:2}b, $\kappa$ is found
to vanish in the plateau regions, as expected for an incompressible crystalline
Fock state, and in overall good agreement with our theoretical results. The
finite values in between result from the small energy gaps between different
Fock states around $\Delta_{\rm c}$, leading to the preparation of compressible
superposition states.

%%%%%%%%%%%%%%%%%%%%%%%%%%%%%%%%%%%%%%%%%%%%%%%%%%%%%%%%%%%%%%%%%%%%%%%%%%%
%          Fig3: Evolution along the path through the phase diagram 1d    %
%%%%%%%%%%%%%%%%%%%%%%%%%%%%%%%%%%%%%%%%%%%%%%%%%%%%%%%%%%%%%%%%%%%%%%%%%%%

% ------- Figure 4 -------
\begin{SCfigure*}
  
\centering
\includegraphics{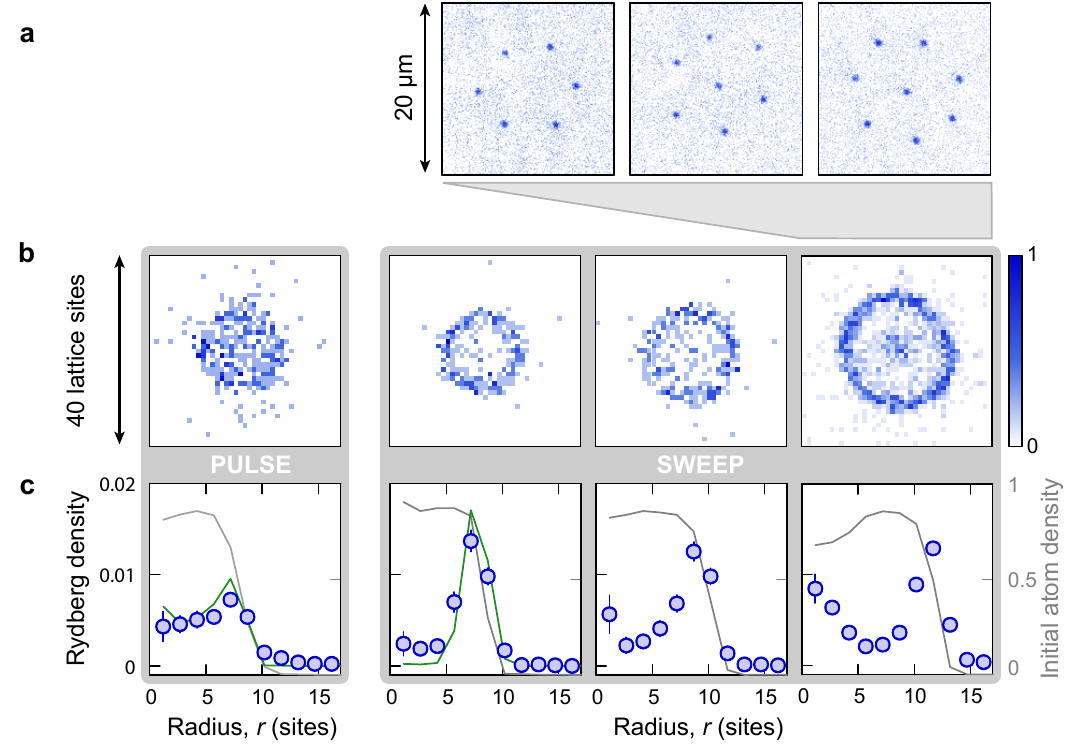} %[width=0.75\textwidth]
\caption{{\bf Dynamical crystallization in two dimensions.}  \textbf{a},
Unprocessed experimental single shot pictures of Rydberg states with $6$, $7$
and $8$ excitations from the rightmost data set. Each blue point corresponds to a single atom. \textbf{b,
c}, Rydberg densities for pulsed excitation (left grey box) and sweeped
excitation with increasing cloud size (right grey box). The pulsed excitation
was done with the same amplitude modulation as for the sweep (Fig.~\ref{edfig:1}b), but the detuning $\Delta$ was held constant (averaged data for $\Delta
= 2\pi\cdot 260\,$kHz and $\Delta = 2\pi\cdot 760\,$kHz is shown). The cloud
radius was \SI{8.2(2)}, \SI{8.3(1)}, \SI{10.0(3)}, \SI{11.8(2)} lattice sites
(left to right). \textbf{b}, Measured two-dimensional distribution of Rydberg
excitations. The colour scale represents the normalized counts per site. 
\textbf{c}, Azimuthally averaged density distribution (probability per site) of
the data shown in \textbf{b} (blue dots) and comparison with theory (green
line). The theoretical calculation was only feasible for small clouds and is based on representative experimental initial atomic samples. Their 
density is shown in grey on the right axis. Error bars s.e.m.} \label{fig:4}

\end{SCfigure*}

\begin{figure}[h!]
\centering
\includegraphics{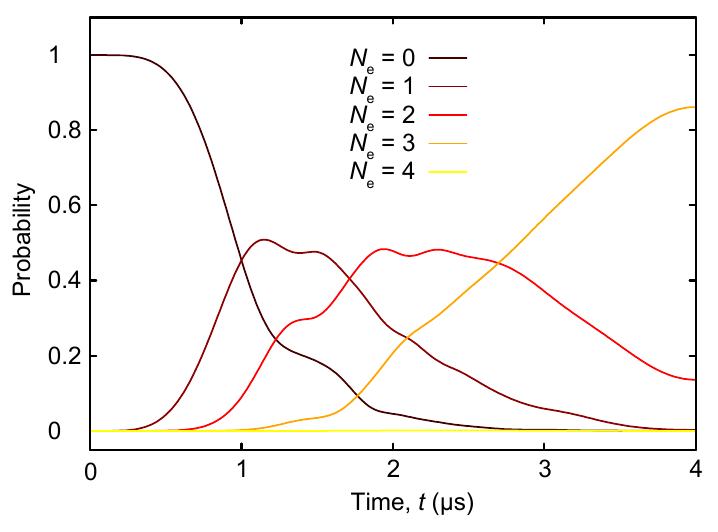}
\caption{{\bf Calculated time-evolution of the Rydberg Fock state population
  during the optimized sweep in 1D.} Calculation of the time-dependent
  occupation of the Fock states for an ideal $3\times23$ chain with
  unity filling.
}\label{edfig:2}
\end{figure}

The preparation involves complex quantum dynamics of the many-body
system. To trace the evolution of the crystallization process along the sweep
trajectory $(\Omega(t),\Delta(t))$ we abruptly switched off the coupling at
different times, thereby projecting the many-body state onto the eigenstates of
the uncoupled system $(\Omega=0)$. For the measurement we chose the optimized sweep for the
$\Ne=3$ Fock state in a system of $3 \times 23$ sites. The path through
the phase diagram is shown in Fig.~\ref{fig:3}a. For each evolution time we
measured the excitation number histogram, from which we extracted the mean
Rydberg number $\langle \hat{\Ne} \rangle$ and its normalized variance $Q =
\frac{\langle \hat{\Ne}^2 \rangle - \langle \hat{\Ne} \rangle^2}{\langle \hat{\Ne} \rangle}-1$
(Fig.~\ref{fig:3}b). During the sweep \Ne increases until we
observe a saturation behaviour which we interpret as the onset of
crystallization (Fig.~\ref{edfig:2}). Simultaneously, the $Q$ factor decreases from the
Poissonian value $Q \approx 0$ to $Q \approx -0.5(1)$, which reflects the approach
to the Fock state. The expected value $Q \approx -1$ is increased to $Q \approx -\alpha$ due to our detection efficiency. 
The measurement of the full counting statistics along the
sweep trajectory allows for a more quantitative comparison with theory
(Fig.~\ref{fig:3}c). However, the finite detection efficiency strongly affects
the observed histograms and leads to a tail of the distributions towards lower
excitation numbers (Methods). Nevertheless, when adapting the theoretical
prediction to take $\alpha=0.62$ into account, we find very good agreement with
the experimental observations.

The high-resolution detection scheme allows for an even more detailed study of
the dynamics via the spatial excitation density, which is largely unaffected by
the finite detection efficiency. At the beginning of the pulse, where the
excitation probability is low, we observe delocalized Rydberg atoms throughout
the cloud (Fig.~\ref{fig:3}d), characteristic for the disordered phase in this
parameter regime. For longer times the excitations start to accumulate at both
ends of the line-shaped cloud and finally crystallize to the expected
triple-peak configuration. The dynamics of this crystallization process
matches well with the theoretical expectations and the slight broadening of the observed
peaks is compatible with the spatial resolution of the detection of one lattice
site~\mycite{Schauss2012}.

%%%%%%%%%%%%%%%%%%%%%%%%%%%%%%%%%%%%%%%%%%%%%%%%%%%%%%%%%%%%%%%%%%%%%%%%%%%
%          Fig4: Adiabatic sweeps in 2d                                    %
%%%%%%%%%%%%%%%%%%%%%%%%%%%%%%%%%%%%%%%%%%%%%%%%%%%%%%%%%%%%%%%%%%%%%%%%%%%

In a different set of experiments we investigated the chirped laser-excitation in
small two-dimensional lattices. We used the spatial light modulator to prepare
disc-shaped clouds with a controlled radius, whose value fluctuated by only one
lattice site (Methods). Here, the dynamical preparation turned out to be
more challenging, since effects of the fluctuating boundary are much more
pronounced in two dimensions than in the effective one-dimensional geometry
discussed above. Nevertheless, a proper frequency chirp of the excitation laser
offers substantial control of the many-body dynamics and the preparation of
energetically low-lying many-body states. This is demonstrated in Fig.~\ref{fig:4} 
where we compare the spatial distribution of Rydberg atoms when
exciting the atoms at a constant detuning to the result of a
chirped excitation from $\Delta_{\rm min}<0$ to $\Delta_{\rm
max}$ (Fig.~\ref{edfig:1}b).  In the former case the excitations are delocalized
across the atomic sample, while in the latter low energy states with localized
excitations are prepared. The initial system size permits to control the
excitation number and attainable spatial structures. With increasing \Ne the
configuration with all Rydberg excitations located along the circumference
becomes energetically unfavourable compared to configurations with an extra
Rydberg atom in the lattice centre. This structural change is directly visible
in the spatial excitation patterns shown in Fig.~\ref{fig:4}. 

%%%%%%%%%%%%%%%%%%%%%%%%%%%%%%%%%%%%%%%%%%%%%%%%%%%%%%%%%%%%%%%%%%%%%%%%%%%
%          Conclusion and Outlook                                         %
%%%%%%%%%%%%%%%%%%%%%%%%%%%%%%%%%%%%%%%%%%%%%%%%%%%%%%%%%%%%%%%%%%%%%%%%%%%

In conclusion, we have prepared and studied the crystalline phase in a Rydberg lattice gas using coherent laser sweeps.
This constitutes the first controlled preparation of
crystalline states in neutral atom systems. Furthermore, these adiabatic
preparation techniques enable the detailed study of the underlying phase diagram, and intriguing phenomena such as
the predicted two-stage melting via a floating crystal
phase~\mycite{Weimer2010, Sela2011}. More generally, our results pave the way towards the
study of quantum correlations and dissipative quantum magnets in long-range
interacting Ising-type spin systems~\mycite{Lesanovsky2010a,Ates2012,Petrosyan2013b}.

	\section*{Acknowledgements}
	We acknowledge funding by MPG, DFG, EU
  (UQUAM, SIQS, ITN-COHERENCE, HAIRS, Marie Curie Fellowship to M.C.).
%%%%%%%%%%%%%%%%%%%%%%%%%%%%%%%%%%%%%%%%%%%%%%%%%%%%%%%%%%%%%%%%%%%%%%%%%%%
%          Bibliography                                       %
%%%%%%%%%%%%%%%%%%%%%%%%%%%%%%%%%%%%%%%%%%%%%%%%%%%%%%%%%%%%%%%%%%%%%%%%%%%

\bibliography{RydbergSweeps_arxiv_final}
  
% \newpage

% figure legends max. around 100 words each.
% figure legends in total less than 500 words.

\section*{Methods}

% ------- Extended Data Figure 1 -------
\begin{figure*}
\centering
\includegraphics{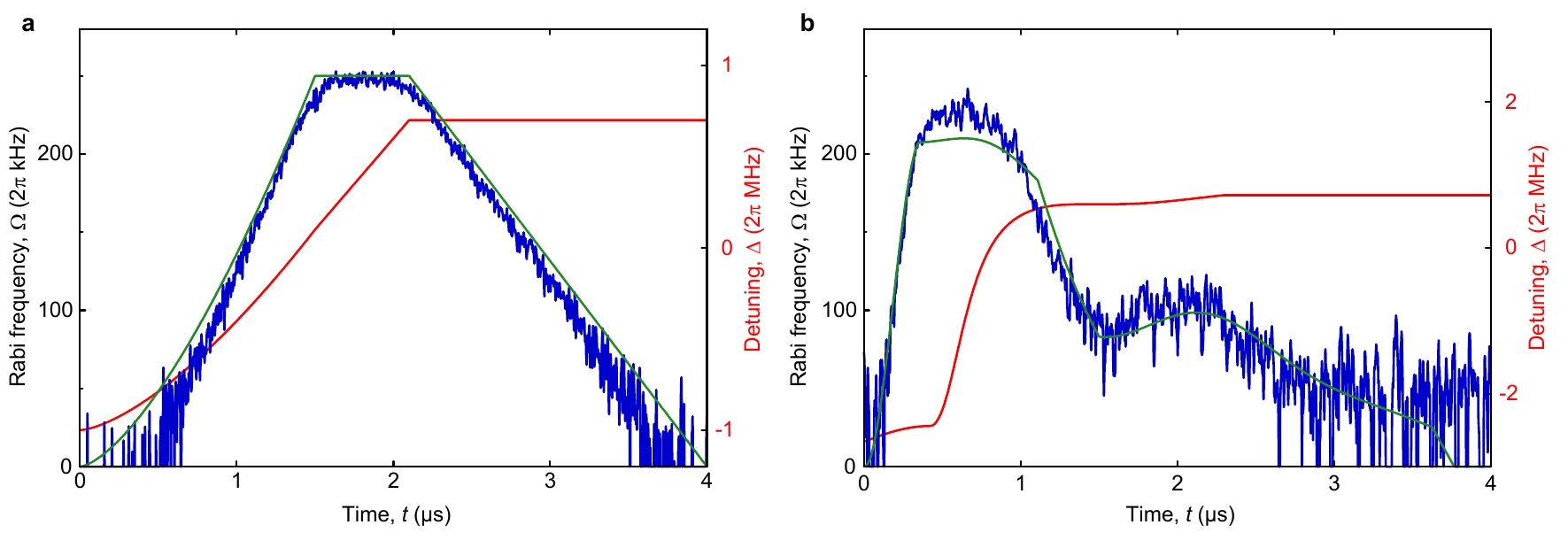}
\caption{{\bf Frequency and amplitude modulation of the sweeps in 1D and
  2D.} Time dependence of Rabi frequency and detuning during the sweep
  used in the experiments with 1D systems (\textbf{a}) and 2D systems (\textbf{b}).
  Blue line, two-photon Rabi frequency obtained by a calibrated photo diode. The noise is due to the low light level on the high-bandwidth photo diode. 
  For the 2D systems the pulse was not optimized using the full procedure discussed in the Methods, since the limiting factor here was the fluctuation of the cloud shape. For the largest 2D systems
  (Fig.~\ref{fig:4}, rightmost column) the Rabi frequency was scaled up by a
  factor of $1.9(1)$ compared to \textbf{b}. Green line, targeted two-photon Rabi frequency. Red
  line, targeted detuning. 
	}\label{edfig:1}
\end{figure*}

%%%%%%%%%%%%%%%%%%%%%%%%%%%%%%%%%%%%%%%%%%%%%%%%%%%%%%%%%%%%%%%%%%%%%%%%%%%
%                  Rydberg excitation and sweep details                   %
%%%%%%%%%%%%%%%%%%%%%%%%%%%%%%%%%%%%%%%%%%%%%%%%%%%%%%%%%%%%%%%%%%%%%%%%%%%
\subsection*{Rydberg excitation}

The excitation path, detection scheme, laser setup and optical lattice
configuration during the Rydberg excitation were explained in a previous
publication~\mycite{Schauss2012}. The beam parameters are summarized in Table~\ref{edtab:1}.

We realized the required frequency and amplitude modulation of the two-photon
coupling by changing the parameters of the red beam (\SI{780}{nm}), while the
parameters for the blue beam (\SI{480}{nm}) were held constant. The fast
modulations were implemented using an acousto-optical modulator in double-pass
configuration at a centre frequency of \SI{350}{MHz}. For the frequency sweep a
synthesizer with large frequency modulation bandwidth was used to drive the
acousto-optical modulator.  The pulse amplitude was shaped using a fast,
calibrated variable attenuator after the synthesizer. Long-term drifts were
minimized by a sample-and-hold intensity stabilization technique. 

We measured the resonance curve of the Rydberg excitation in a very dilute
cloud to avoid interaction induced broadening and shifts. Over consecutive
days the line centre was reproducible  within about \SI{200}{kHz}. The
linewidth of the Rydberg lasers was \SI{50}{kHz} for the red laser and
\SI{20}{kHz} for the blue laser. To measure the Rabi frequency $\Omega$ of the
two-photon coupling we prepared a single atom using our addressing technique
and observed its Rabi oscillations.

The use of the red laser to implement the amplitude modulation lead to
a time-dependent Stark shift on the ground state, which caused an additional
detuning of the transition to the Rydberg state. This effect had to be
compensated during the sweep, as it can easily exceed the two-photon Rabi
frequency. The Stark shifts on the intermediate and Rydberg states were
negligible, the former due to the vanishing influence of the level shift on the
excitation scheme, the latter due to the weak coupling. The amplitude and
frequency modulation throughout the sweeps are shown in Fig.~\ref{edfig:1}.

The relatively small beam waist of the excitation beams
(Table~\ref{edtab:1}) lead to a spatial variation of the Rabi
frequency $\Omega$ across the system. For most of the measurements the
difference in coupling strength was less than $\SI{30}{\percent}$, but for the
largest one-dimensional systems with $\ell>26$  up to $\SI{40}{\percent}$. 

%%%%%%%%%%%%%%%%%%%%%%%%%%%%%%%%%%%%%%%%%%%%%%%%%%%%%%%%%%%%%%%%%%%%%%%%%%%
%                  Rydberg detection scheme and push out                  %
%%%%%%%%%%%%%%%%%%%%%%%%%%%%%%%%%%%%%%%%%%%%%%%%%%%%%%%%%%%%%%%%%%%%%%%%%%%

% \newpage
% ------- Extended Data Figure 2 -------

% \newpage
% ------- Extended Data Figure 3 -------
\begin{figure}[h!]
\centering
\includegraphics{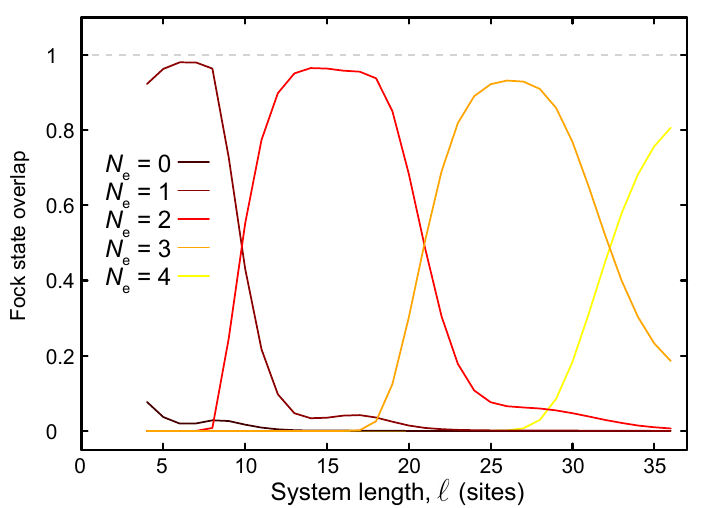}
\caption{{\bf Calculated dependence of the Rydberg Fock state population after the sweep on the system length.} Calculation of the Fock state overlap of the state after the sweep for a $3\times\ell$ system taking into account length fluctuations and a filling of $0.8\,$atoms/site of the initial atom configurations.
}\label{edfig:3}
\end{figure}

\subsection*{Rydberg detection}

For detection of the Rydberg atoms we removed the ground state atoms shortly
after the excitation, followed by the de-excitation of the Rydberg atoms using
resonant coupling to the short-lived $5P_{3/2}$ state. Finally, we detected the
remaining atoms via fluorescence imaging in the optical lattice. The ground
state push-out had to be as short as possible, to allow for high detection
efficiency and high spatial resolution. At the same time a very high pushing
efficiency was required. We optimized the ground state removal under these
restrictions to a short pulse of \SI{6}{\micro\second} in total. A re-pumping
laser on the $5S_{1/2},\,\ket{F,m_F}=\ket{1,-1}\leftrightarrow
5P_{3/2},\,\ket{2,-2}$  transition was switched on for the full
\SI{6}{\micro\second} while the push beam
($5S_{1/2},\,\ket{2,-2}\leftrightarrow 5P_{3/2},\,\ket{3,-3}$) was
delayed by \SI{2}{\micro\second}. The Rabi frequencies on both transitions were
approximately $2\pi\cdot\SI{70}{MHz}$. We used a relatively high re-pump power to
include also the other Zeeman sub-states of the $5S_{1/2},\,\ket{F=1}$
manifold, which were split by $\Delta m_F\cdot\SI{21}{MHz}$ for our magnetic
field of \SI{30}{G}. The repump beam is required to remove the very small fraction of atoms in $\ket{F=1}$ states, created by
imperfect state preparation, scattering of the red Rydberg beam and non-perfect
polarization of the push-beam. This scheme allows to reach a push-out efficiency of \SI{99.9}{\percent} within the limited time.

\subsection*{Limits for the detection efficiency}

Our detection method allows for high-resolution imaging of the Rydberg atoms.
However, this requires fast de-pumping of the Rydberg atoms into the ground
state. Initially the ground state atoms were prepared in the lowest band of the
optical lattice. The excitation to the slightly anti-trapped Rydberg state
caused an expansion of the centre of mass wave function during the push-out
time. The instantaneous de-pumping to the ground state then lead to a
projection of the expanded and, due to the Rydberg-Rydberg interaction,
slightly shifted wave function onto the lattice eigenstates, such that several
bands in the lattice were populated. Before switching on the optical molasses
for the fluorescence imaging we had to wait for \SI{57}{\milli\second} to make sure that the absolute value of the magnetic field is
below \SI{100}{\milli G}, which is required for the molasses to work properly. During this time, the population in higher bands
tunnelled through the lattice. Most of these atoms were not detected and the
rest caused a flat background in the images.

The detection efficiency of $\alpha=0.62(5)$ is determined from the data shown in Fig.~\ref{fig:2}a. The determined value is consistent with independent measurements in Fig.~\ref{fig:3}c and Fig.~\ref{fig:4}.
  
%%%%%%%%%%%%%%%%%%%%%%%%%%%%%%%%%%%%%%%%%%%%%%%%%%%%%%%%%%%%%%%%%%%%%%%%%%%
%                        Initial state preparation                        %
%%%%%%%%%%%%%%%%%%%%%%%%%%%%%%%%%%%%%%%%%%%%%%%%%%%%%%%%%%%%%%%%%%%%%%%%%%%

\subsection*{Initial atomic sample preparation}
  
After cooling the rubidium-87 atoms the experiment began with a
2D-Bose-Einstein condensate in the hyperfine state $\ket{1,-1}$ trapped in the
vertical lattice. We used a de-confining beam at \SI{670}{\nano\meter} to
reduce the radial trap frequencies. To reduce interferences on this beam, it
was generated using a super-luminescent diode (linewidth \SI{\approx
5}{\nano\meter}) and then amplified in two consecutive tapered amplifiers.

Next, we adiabatically switched on the horizontal lattices and created an
atomic Mott-insulator~\mycite{Greiner2002}. We made sure that the centre of the system was in the
unity filling Mott lobe by adjusting the atom number. This allowed to create
unity filling regions of a diameter of up to \SI{15}{\micro\meter}. Only for
very large systems (c.f. Fig.~\ref{fig:2}a, $\ell > 27$ and Fig.~\ref{fig:4}c, rightmost
panel) we had to allow for double occupancy in the central part. 

A spatial light modulator was used to project a line- or disc-shaped pattern
with adjustable size onto the atoms~\mycite{Fukuhara2013a}. Intensity,
polarization and detuning of this light pattern were chosen to create a
differential light shift on the transition \mbox{$\ket{1,-1}\leftrightarrow\ket{2,-2}$} of approximately \SI{50}{kHz}, but nearly no absolute shift of the
\mbox{$\ket{1,-1}$} state. Next, the atoms on non-shifted sites were transferred to
the $\ket{2,-2}$ state using a microwave sweep and subsequently removed by
resonant laser light on the \mbox{5S$_{1/2}\,,\ket{2,-2}\leftrightarrow
5P_{3/2}\,,\ket{3,-3}$} transition. The final step was to transfer the remaining
atoms via a microwave sweep to the $\ket{2,-2}$ state. The average filling in
the remaining part of the system was $0.8\,$atoms/site.

%%%%%%%%%%%%%%%%%%%%%%%%%%%%%%%%%%%%%%%%%%%%%%%%%%%%%%%%%%%%%%%%%%%%%%%%%%%
%            Theory: Optimization techniques and adiabaticity             %
%%%%%%%%%%%%%%%%%%%%%%%%%%%%%%%%%%%%%%%%%%%%%%%%%%%%%%%%%%%%%%%%%%%%%%%%%%%
\subsection*{Numerical pulse optimization} %{Optimization of the adiabatic pulses}
Our theoretical calculations are based on a numerical solution of the Schr\"odinger
equation for the Hamiltonian (1) in a truncated Hilbert space. To this end, we expand the wave 
function $\ket{\psi}$ of the \Nat-atom system in terms of Fock-states, which are eigenstates of the classical Hamiltonian
(i.e. for $\Omega=0$)
\begin{widetext}
\begin{equation}
    \ket{\psi} = c^{(0)}\ket{0} + \sum_{{\vec i}_1} c^{(1)}_{{\vec i}_1}\ket{{\vec i}_1} +
    \sum_{{\vec i}_1, {\vec i}_2} c^{(2)}_{{\vec i}_1,{\vec i}_2}\ket{{\vec i}_1, {\vec i}_2} +
    \ldots + \sum_{\mathclap{{\vec i}_1, \ldots, {\vec i}_{\Nat}}} c^{(\Nat)}_{{\vec i}_1, \ldots,
   {\vec i}_{N_{\rm at}}}\ket{{\vec i}_1, \ldots, {\vec i}_{\Nat}} \;,
\end{equation}
\end{widetext}
where $\ket{{\vec i}_1, \ldots, {\vec i}_{N}}$ corresponds to a state with
$N$ Rydberg excitations located at lattice sites ${\vec i}_1$ to ${\vec
i}_N$, and $c^{(N)}_{{\vec i}_1, \ldots, {\vec i}_{N}}$ denotes the
corresponding time-dependent amplitude. In order to truncate the otherwise exponentially large Hilbert space 
we only include excitation numbers of $N\leq N_{\rm c}$ and introduce a cut-off distance $R_{\rm c}$, 
discarding all states that contain Rydberg excitations closer than $R_{\rm c}$ \mycite{Schauss2012}. 
The presented results were obtained for $N_{\rm c}=5$ and we verified directly that the inclusion
of states with $N=6,7$ did not contribute to the many-body dynamics. In addition $R_{\rm c}$ was reduced 
progressively until convergence of the simulated dynamics was achieved~\mycite{Schauss2012}. 
	
In order to optimize the excitation pulse we monitor the fidelity
\begin{equation}
\mathcal{F}( \ket{\psi(t)}, \ket{\psi_{\text{GS}}(t)} ) = \abs{ \langle \psi_{\text{GS}}(t)|\psi(t)\rangle }^2\;,
\end{equation}
i.e., the overlap between the time-evolving wave function $\ket{\psi(t)}$ and the many-body ground
state  $\ket{\psi_{\text{GS}}(t)}$ at time $t$ determined by the actual set of laser parameters $\Omega(t)$ 
and $\Delta(t)$. First we choose a given trajectory $(\Omega(t),\Delta(t))$ that connects the initial ground state 
with all atoms de-excited with the targeted crystalline ground state. The actual path chosen for our experiments 
is shown in Fig.~\ref{fig:3}a. Next we discretize the path into a large number of sampling points, forming equal linear 
segments, along which we propagate the Schr\"odinger equation as described above. A suitable pulse for 
high-fidelity Fock-state generation was then obtained by optimizing the local 
ramp speed for each segment with respect to the decrease of $\mathcal{F}(t)$ between successive 
sampling points, with the constraint $T=\SI{4}{\micro\second}$ for the total pulse duration, $T$. As described in the main text, 
non-adiabatic transitions between closely lying states with equal $N$ but finite dislocations are unavoidable in 
the final stage of the excitation pulse. Hence, for the chosen path, we stop the optimization and fixed the ramp 
speed once the Rabi frequency drops below \SI{60}{kHz}. Since the experimentally prepared atomic lattice has a 
finite filling fraction of $\approx0.8$, this procedure is repeated for several random configurations producing 
slightly different optimal pulses. The pulse shown in Figure~\ref{edfig:1}a was 
constructed as a simple and simultaneously good compromise between those numerically obtained pulses for $\ell=23$.
As demonstrated in Fig.~\ref{edfig:3}, this pulse indeed yields a high final Fock-state fidelity of $0.98$, $0.96$ and 
$0.93$ for $\Ne=1$, $\Ne=2$ and $\Ne=3$ excitations, respectively. In our experiments, this pulse was used to obtain the 
Rydberg blockade staircase shown in Fig.~\ref{fig:2}.

%%%%%%%%%%%%%%%%%%%%%%%%%%%%%%%%%%%%%%%%%%%%%%%%%%%%%%%%%%%%%%%%%%%%%%%%%%%
%             Theory: Finite size scaling and compressibility             %
%%%%%%%%%%%%%%%%%%%%%%%%%%%%%%%%%%%%%%%%%%%%%%%%%%%%%%%%%%%%%%%%%%%%%%%%%%%
\subsection*{Estimating the compressibility}

In the dilute limit, $\Ne \ll \ell$, the Hamiltonian Eq.~(1) for a 1D chain can
be reformulated in a continuous form. To this end, we scale lengths and
energies by $L=\ell \alat$ and $C_6/L^6$, and introduce dimensionless operators 

\begin{equation} \label{eq:psi}
\begin{split}
\hat{\psi}(x_i)&=\sqrt{\frac{L}{\alat}}\ket{g^{( i)}}\bra{e^{( i)}}=\sqrt{\ell}\ket{g^{( i)}}\bra{e^{( i)}}\;,\\
\hat{\psi}^{\dagger}(x_i)&=\sqrt{\ell}\ket{e^{( i)}}\bra{g^{( i)}}\;,
% \\
% \text{~}
\end{split}
\end{equation}

where $x_i=i\alat/L$. Taking the continuum limit, this permits to write the lattice Hamiltonian as
\begin{widetext}
\begin{equation}\label{eq:H_cont}
  \hat{H}=B_\perp\int\limits_0^1(\hat{\psi}^{\dagger}(x)+\hat{\psi}(x))\,{\rm d}x + 
          B_\parallel\int\limits_0^1\hat{\psi}^{\dagger}(x)\hat{\psi}(x)\,{\rm d}x + 
          \frac{1}{2}\int\limits_0^1\!\!\!\int\limits_0^1\frac{\hat{\psi}^{\dagger}(x)\hat{\psi}^{\dagger}(x^{\prime})\hat{\psi}(x^{\prime})\hat{\psi}(x)}{|x-x^{\prime}|^6}\,{\rm d}x\,{\rm d}x^{\prime}
\end{equation}
\end{widetext}
with the dimensionless, effective magnetic fields
$B_\perp=\sqrt{\ell}\frac{\hbar\Omega}{2}\frac{L^6}{C_6}$ and
$B_\parallel=-\hbar\Delta\frac{L^6}{C_6}$. The quantum fields
$\hat{\psi}^{\dagger}$ and $\hat{\psi}$ describe the creation and annihilation
of hard-core bosons, where the hard-core constraint is naturally ensured by the
Rydberg-Rydberg atom interaction. The excitation number, and, hence, the
density increases linearly with $L$, such that $L\rightarrow\infty$ defines the
thermodynamic limit. 

Consequently the excitation number $\Ne=\Ne(B_\parallel,B_\perp)$ depends on
only two parameters and we can re-express the compressibility
$\kappa=\frac{\partial \Ne}{\partial \Delta}$ in terms of the derivative of
$\Ne$ with respect to $\ell$. Neglecting the weak dependence on $B_\perp$, thus, yields
\begin{equation}\label{eq:kappa}
% \frac{\partial \Ne}{\partial \ell} \frac{\partial \ell}{\partial \Delta}=
\kappa=\frac{\ell}{6\Delta}\frac{\partial \Ne}{\partial \ell}\;,
\end{equation}
which we used to calculate the compressibility $\kappa$ from the measured
dipole-blockade staircase shown in Fig.~\ref{fig:2}. The required $\ell$-derivative was obtained from 
the experimental data, $\Ne^{(i)}$ and $\ell^{(i)}$, by using a second-order symmetric formula~
\mycite{Fornberg1988}: $\left.\frac{\partial\Ne}{\partial\ell}\right|_{\ell=\bar{\ell}^{(i)}} \approx\frac{\Ne^{(i+1)}-\Ne^{(i-1)}}{\ell^{(i+1)}-\ell^{(i-1)}}$ with $\bar{\ell}^{(i)} = \frac{1}{2}(\ell^{(i+1)}+\ell^{(i-1)})$.

% ------- Extended Data Table 1 -------
% tables have to be sans-serif corresponding to nature guide.
\begin{table*}
\centering%
% \sffamily%
% \sisetup{detect-family}%
\begin{tabular}{llrrr}
  
  \toprule & & \multicolumn{1}{c}{red pulse 1D} & \multicolumn{1}{c}{red pulse 2D} & \multicolumn{1}{c}{blue} \\
  \midrule
  max. Rabi freq 2photon & $(2\pi\cdot\si{kHz})$ & \tablenum[table-format=3]{250+-25}  & \tablenum[table-format=3]{420+-42} & \multicolumn{1}{c}{-} \\
  max. Rabi frequency & $(2\pi\cdot\si{MHz})$ & \tablenum[table-format=3]{29+-2} & \tablenum[table-format=3]{49+-3} &  \tablenum[table-format=3.1(1)e1]{13+-2} \\
  max. Intensity & $(\si{mW/cm^2})$ & \tablenum[table-format=3]{76+-5} & \tablenum[table-format=3]{215+-14} &  \tablenum[table-format=3.1(1)e1]{1.4+-0.3 e 6} \\
  beam waist & $(\si{\micro\metre})$ & \tablenum[table-format=3]{44 +- 2} & \tablenum[table-format=3]{44 +- 2} & \tablenum[table-format=3.1(1)e1]{17 +- 5} \\
  max. light shift & $(2\pi\cdot \si{kHz})$ & \tablenum[table-format=3]{236+-35} & \tablenum[table-format=3]{670+-100} & \tablenum[table-format=3.1(1)e1]{57+-9} \\
  \bottomrule
\end{tabular}
\caption{{\bf Parameters of the Rydberg laser beams.} For the red laser in the 1D configuration (first column), and 2D (second column). The blue laser (third column) has the same parameters in 1D and 2D. Errors, s.d.}
\label{edtab:1}
\end{table*}

In order to verify the quality of the employed approximations, we also
determined $\kappa$ numerically directly from its definition. To this end, we
recorded the excitation number staircase using the excitation pulse
$(\Omega(t),\Delta(t))$ shown in the Figure~\ref{edfig:1}a and for a 
reference pulse with a small global frequency shift $\delta$, i.e. for
$(\Omega(t),\Delta(t)+\delta)$. Within the studied range $|\delta/(2\pi)|<
\SI{10}{kHz}$ the number of excitations \Ne was found to vary linearly with
$\delta$ such that the compressibility can be evaluated directly via
$\kappa=(\Ne(\delta)-\Ne(\delta=0))/\delta$. This compressibility
$\kappa$ is shown in Fig.~\ref{fig:2}b and agrees well with the experimental
results obtained from equation~\eqref{eq:kappa}.

\end{document}